\documentclass[lettersize,journal]{IEEEtran}
\usepackage{graphicx}
\usepackage{cite}
\usepackage{amsmath,amssymb,amsfonts}
\usepackage{xcolor}
\usepackage{textcomp}
\usepackage{hyperref}
\usepackage{multicol}
\usepackage{bm}
\usepackage{setspace}
\usepackage{multirow}
\usepackage{epstopdf}
\usepackage{comment}
\usepackage[font=footnotesize]{caption}
\usepackage{array}
\usepackage{subcaption}
\usepackage{stfloats}

\usepackage{steinmetz}
\usepackage{algpseudocode}
\usepackage{algorithm}
\usepackage[capitalise]{cleveref}
\usepackage{pifont}
\usepackage{verbatim}
\usepackage{url}
\usepackage{tikz}
\usepackage{tabularx}

\makeatletter
\newcommand{\thickhline}{%
    \noalign {\ifnum 0=`}\fi \hrule height 0.8pt
    \futurelet \reserved@a \@xhline
}
\newcolumntype{"}{@{\hskip\tabcolsep\vrule width 1pt\hskip\tabcolsep}}
\makeatother

\begin{document}

\title{Replicating Real-World 23-Hz Oscillations Caused by Large Electronic Loads}

\author{
Lingling~Fan,~\IEEEmembership{Fellow,~IEEE}, Ali~Yazdanpanah,~\IEEEmembership{Member,~IEEE}, Yunzhi~Cheng,~\IEEEmembership{Senior Member,~IEEE}, Zhixin~Miao,~\IEEEmembership{Senior Member,~IEEE}, Farshid~Salehi,~\IEEEmembership{Member,~IEEE}, Patrick  Gravois,~\IEEEmembership{Senior Member,~IEEE},  Shun-Hsien (Fred) Huang,~\IEEEmembership{Member,~IEEE} \\
IEEE PES IBR/IBL SSO Task Force
 \thanks{
L. Fan and Z. Miao 
are with the Department of Electrical Engineering, University of South Florida,
Tampa,
FL, 33620. E-mail: \texttt{linglingfan@usf.edu}. A. Yazdanpanah, Y. Cheng, P. Gravios, and S. Huang are with ERCOT. 
F. Salehi is with DNV.
}%
}
\maketitle

\begin{abstract}
In 2024, Texas operators observed 23-Hz oscillations in real power measurements close to a large electronic load (LEL). Oscillations emerged when the load’s power consumption reached approximately 320 MW level and subsided as the active power demand decreased. The paper aims to analyze the event and reproduce the oscillations using electromagnetic transient (EMT) simulations.
In the first stage, a representative feedback system is developed, and frequency-domain analysis is conducted to examine the phenomenon and identify its key influencing factors. Next, detailed EMT simulations are performed to further validate the proposed analytical approach. The results show that the feedback system effectively captures and characterizes the critical features of the 23-Hz oscillation incident. In addition, the EMT simulations successfully reproduce the real-world event, with the simulated results closely matching the fault recorder data.

 
\end{abstract}


\section{Introduction}

\IEEEPARstart{I}{n} recent years, the share of large electronic loads (LELs) interfaced with the grid through power electronic converters (e.g., data center, crypto mining) has increased significantly. The substantial size, geographic concentration, and power-electronic-based characteristics of these loads have introduced new reliability challenges for system operators. For instance, ERCOT has observed that such loads are highly sensitive to voltage variations and tend to reduce consumption immediately during moderate voltage disturbances \cite{Gravois_blog}. Conversely, they are capable of rapidly increasing their power demand. A team from the leading IT industry has published paper on data center power consumption pattern \cite{choukse2025power}, demonstrating such a fast power varying pattern.These abrupt load fluctuations can negatively impact system frequency and voltage stability. To address these challenges, the power industry is actively collaborating to develop necessary requirements such as voltage ride-through and ramp rate limits and to incorporate these behaviors into system models that enable more accurate and comprehensive reliability assessments.


In addition to traditional stability concerns, LELs have caused oscillations in power grids. The oscillations are attributed to their inverter-based interface. In 2017, Meta’s data centers experienced oscillations of $49$ Hz and $71$ Hz in three-phase currents and voltages \cite{sun2022data}. An investigation in  \cite{sun2022data} identified the DC-link voltage control in the server power supply unit (PSU) as a major contributor. Impedance-based analysis revealed that the DC-link voltage control parameters significantly affect the PSU’s impedance frequency response, leading to the observed oscillations. 

More recently, Dominion Energy observed $14.7$-Hz oscillations with a 4\% peak-to-peak magnitude  in a $115$-kV substation's voltage measurements in a data center region \cite{mishra2025understanding}. The oscillations emerged when the four hydro units closeby the region ramped down power. Measurement-based analysis shows that the oscillations are attributed to the data center's uninterruptible power supply (UPS) systems. 

In October 2024, Texas operators observed $23$-Hz oscillations in real power measurements \cite{ERCOT_largeload}. Subsequent investigation identified a LEL as the source.  When the load’s power consumption reached approximately $320$ MW level, oscillations appeared and when the power consumption level was reduced, oscillations disappeared. Eventually, after firmware upgrade in the LEL, oscillations no longer occurred.


The goal of this paper is to analyze the $23$-Hz oscillation event. Although modern tools (e.g.,  high-fidelity EMT simulation) are available, no dynamic model for the oscillating load has been submitted by the load entity and the associated transmission service provider (TSP) due to limited operational experience and lack of industry accepted dynamic model for LELs. A LEL model in the EMT platform recently developed jointly by Texas A\&M University and ERCOT was used in this paper to analyze and reproduce the event \cite{ERCOT_PSCAD} \footnote{The representative model developed in this study is for research purposes only and does not replace the required model submission by the load entity or Transmission Service Provider (TSP).}  On the other hand, it remains a challenging task to tune the parameters of the LEL model and demonstrate $23$-Hz oscillations. 



While the impedance-based analysis adopted in \cite{sun2022data} can tell stability vs. instability, this method relies on the measured frequency responses of the impedance of a PSU, which is essentially a black-box input/output model or a transfer function (matrix). It is challenging to use such an analysis approach to tell exactly how the internal control inside PSUs can interact with the grid, and further lead to oscillations in the certain frequency range. 

\emph{\bf Our contributions:} In this research, we tackle root cause analysis of a real-world event by examining the field data and further developing a customized feedback system. Such a feedback system, with key control units explicitly modeled, can elucidate the mechanism of the $23$-Hz oscillations. The analysis results obtained based on this feedback system were then used to tune an EMT testbed to reproduce $23$-Hz oscillations. 

\emph{Application scope:} It has to be noted that the root cause analysis provided in this paper is for this particular $23$-Hz event in Texas, rather for any data center related oscillation events, since not all events share the similar characteristics. For instance, the root cause analysis results in this paper are not applicable to the $14.7$-Hz event observed in Dominion Energy  \cite{mishra2025understanding}, as that event is associated with voltage, instead of real power. 

The rest of the paper is organized as follows. Section II presents the real-world $23$-Hz oscillation event and the initial analysis.  How to construct the customized feedback system is presented in Section III, along with the analysis results to demonstrate $23$-Hz oscillations. In Section IV, we present the EMT testbed circuit diagram, device-level details, and converter control details. We also present the major simulation results to demonstrate that a high-power consumption level leads to $23$-Hz oscillations. What's more, the time-domain simulation results have been compared with the real-world digital fault record data to show almost exact match in oscillation frequency and peak-peak magnitude. Conclusions are presented in Section V.

\section{The real-world 23-Hz oscillation event and initial analysis}

\begin{figure*}[!ht]
\centering
\includegraphics[width=6.0in]{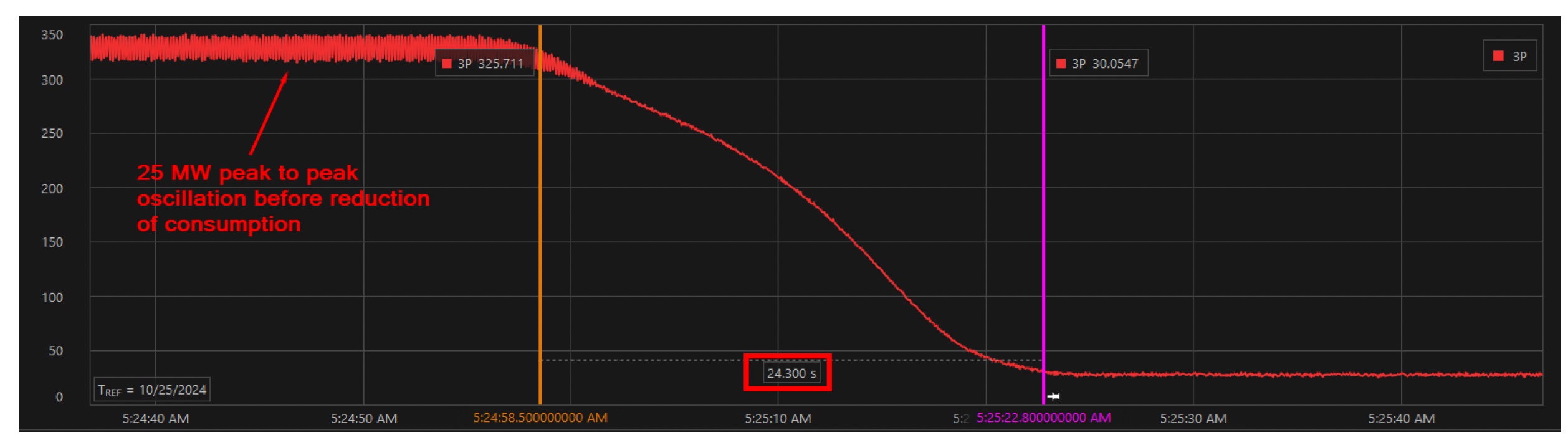}
\caption{The 23-Hz oscillation event in Texas on Oct. 25, 2024. This plot shows real power measurement to the load. }
\label{fig:event10252024}
\end{figure*}

On October 25, {2024}, ERCOT observed a LEL dropped around 300 MW over 24 seconds within a single telemetry scan. Over the previous week, this load had increased consumption to around 330 MW. PMU data showed no fault preceding and the oscillation magnitude preceding reduction was 25 MW peak to peak and 7.5 Hz. It was later determined that the true oscillation mode was 23 Hz from higher resolution data. Fig. \ref{fig:event10252024} shows the real power measurement of this 23-Hz oscillation event as well as power reduction. 

\begin{figure}[!ht]
\centering
\includegraphics[width=3.5in]{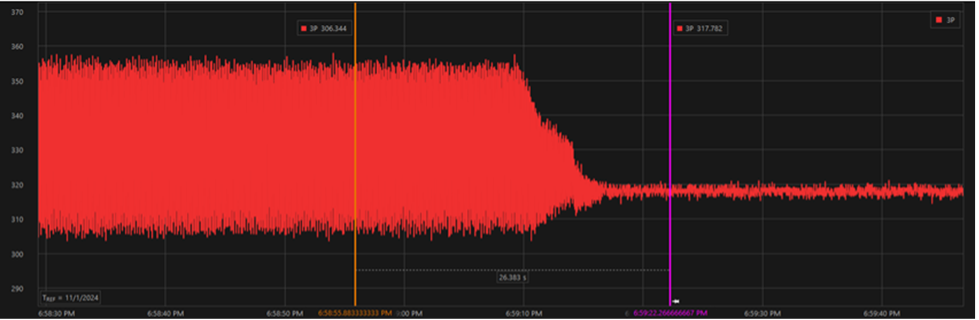}
\caption{This plot shows real power measurement to the load on November 1, 2024. }
\label{fig:fig1}
\end{figure}

Follow-up tests conducted by the load owner/operator show that the load could be reduced to the level that oscillations are no longer present. Fig. \ref{fig:fig1} presents the power measurement data from those tests on Nov. 1, 2024. {The oscillation frequency is 23 Hz and the peak-peak magnitude is 50 MW.} This set of data will be used to compare with the EMT simulation results in Section IV. 

\subsection{Critical features}
There are several critical features of the $23$-Hz oscillations observed in Texas. First, the oscillations occurred when the load's power consumption reached above 320 MW. {This load was connected to a strong grid and there was no transmission outage.} When the power consumption from the load reduced, oscillations were mitigated. In many of the real-world oscillation events associated with inverter-based resources (IBRs) \cite{cheng2023real}, increasing power generation level and reducing grid strength are two major causes that triggered oscillations. For example, ERCOT's 4-Hz oscillations in wind farms \cite{huang2012voltage} were triggered due to a line tripping event\textemdash which made the short circuit ratio (SCR) reduce from $4$ to $2$. Additionally, the oscillations became more severe when the real power exporting level was full. The $23$-Hz oscillation event has something similar to those IBR oscillation events.  

Second, the oscillations were observed in real power. This means that the most observable measurement for the oscillations is real power. In contrast, the $14.7$-Hz oscillations observed in Dominion Energy's data center region were most observable in the voltage measurement. Similarly, the $4$-Hz oscillations in the wind farms were most evident in the voltage measurements, and they can be mitigated by adjusting the power plant voltage controller \cite{huang2012voltage}. This feature helps us exclude one cause of oscillations, i.e., interactions of reactive power and voltage and improper parameter setting of the power plant level voltage control.

\begin{figure}[!ht]
\begin{center}
\includegraphics[width=3.0in]{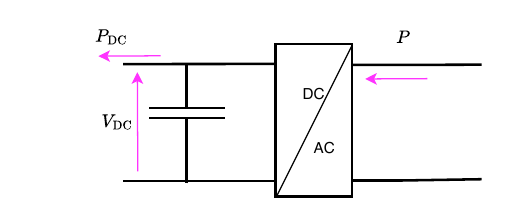}
\end{center}
\caption{DC-link capacitor.}
\label{fig:dccircuit}
\end{figure}

These two features are also the characteristics of the $49$-Hz and $71$-Hz oscillations observed in a Meta Data center. While  \cite{sun2022data} has not mentioned the observed oscillation frequency in real power, we may infer that in the real power measurements, the oscillation frequency is $11$ Hz. Furthermore, \cite{sun2022data} shows that the data center employs many PSUs to convert AC electricity to DC electricity. In these PSUs, power factor corrections (PFCs) have been used to ensure that the AC current flowing to a PSU is in phase with the AC voltage \cite{mao1997review}. It can be seen that PFCs serve the purpose of synchronizing the load current with the grid voltage. Therefore, PFCs are the synchronizing units for LELs.  In addition, the magnitude of the current is determined by a DC-link voltage controller. When the real power absorbed from the AC grid increases, the DC-link voltage also increases. According to Fig. \ref{fig:dccircuit}, the DC-link capacitor's dynamic is described as follows:
\begin{align}
C_{\rm dc} \frac{dV_{\rm dc}}{dt}. V_{\rm dc} = P -P _{\rm dc} 
\end{align}

This equation has been discussed in \cite{fan2018wind} and its per unitized equation is expressed as follows:
\begin{align}
 \tau_{\rm DC} \frac{dV^2_{\rm dc}}{dt} = P -P _{\rm dc} 
\end{align}
where $\tau_{\rm DC} = \frac{\frac{1}{2}C_{\rm dc} V^2_{\rm dc}}{P_{\rm base}}$.
At the nominal operating condition, in per unit (pu), $V_{\rm dc} \approx 1$. Therefore, the above equation further leads to the following in pu:
\begin{align}
&2\tau_{\rm DC} \frac{dV_{\rm dc}}{dt}   \approx P -P _{\rm dc} \\
\text{Or: } &\frac{\Delta V_{\rm dc}}{\Delta P} = \frac{1}{2\tau_{\rm DC} s}
\end{align}
where $P$ refers to the AC side power injection to the load.  

The DC-link voltage controller takes the error between the DC-link voltage reference and the measured voltage 
and passes the error through a proportional and integral (PI) controller to generate the current magnitude. 

An increment in the DC power requirement will lead to a decrease in the DC-link voltage level. In turn, the DC-link voltage control tries to increase the real current injection to the load, which can help increase the real power from the AC side, therefore, balancing the DC power requirement.

\section{Feedback system construction and quantitative analysis}
Based on the above reasoning, a feedback system can be constructed. The most difficult part of feedback system construction is to incorporate the PFC control where the rectifier current or AC current has to be dealt with. If those currents are used, it is not possible to have a straightforward feedback system suitable for linear analysis. Based on the fact that a PFC ensures that the AC current and the AC voltage are in phase, it can be reasoned that the AC current has only  the real current component. Therefore, it is most efficient to adopt a rotating $dq$ frame, which is aligned with the AC voltage space vector at steady state. The DC-link voltage control influences the real current $i_d$. 

When this DC-link voltage control is designed, the grid voltage is assumed to be very stiff or remains at $1$ pu. And $i^*_d$ is the same as $i_d$ or $P$ when the voltage is assumed to be $1$ pu.  
\begin{align}
\left(K_p + \frac{K_i}{s} \right)(\Delta V^*_{\rm dc} -\Delta V_{\rm dc}) = \Delta i^*_d =\Delta P
\end{align}

The closed-loop DC-link voltage control system can be found by incorporating the DC-link capacitor dynamic equation. 
\begin{align}
\Delta V_{\rm dc} = \frac{1}{2\tau_{\rm DC} s}\Delta P = \frac{1}{2\tau_{\rm DC} s}\left(K_p + \frac{K_i}{s}\right)(\Delta V^*_{\rm dc} -\Delta V_{\rm dc}) 
\end{align}
\begin{align}
G_{\rm DVC} &=\frac{\Delta V_{\rm dc}}{\Delta V^*_{\rm dc}}= \frac{\left(K_p + \frac{K_i}{s}\right)\frac{1}{2\tau_{\rm DC} s}}{1+ \left(K_p + \frac{K_i}{s}\right)\frac{1}{2\tau_{\rm DC} s}} \notag \\
&= \frac{K_ps + K_i}{2\tau_{\rm DC}s^2+K_p s + K_i}
\label{eq:G_DVC}
\end{align}

\begin{figure}[htbp]
\begin{center}
\includegraphics[width=3.3in]{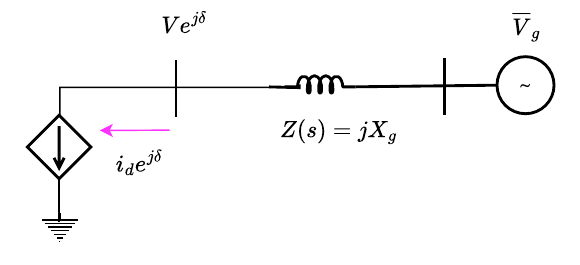}
\caption{A simplified circuit to represent a LEL.}
\label{fig:circuit}
\end{center}
\vspace{-0.15in}
\end{figure}

In real operation, the AC voltage is not stiff and it is influenced by the power drawn from the load. A simple AC circuit is shown in Fig. \ref{fig:circuit}, where a LEL is treated as a controllable current source connected to an interconnection bus. The grid is modeled as a Th\'evenin equivalent with a constant voltage source $\overline{V}_g$ behind an inductive reactance $X_g$.

It can be seen that based on Kirchhoff Voltage Law (KVL), the following equation exists:
\begin{align}
Ve^{j\delta} = \overline{V}_g -jX_g\cdot i_d e^{j\delta}. 
\end{align}

\begin{figure}[!ht]
\begin{center}
\includegraphics[width=3.5in]{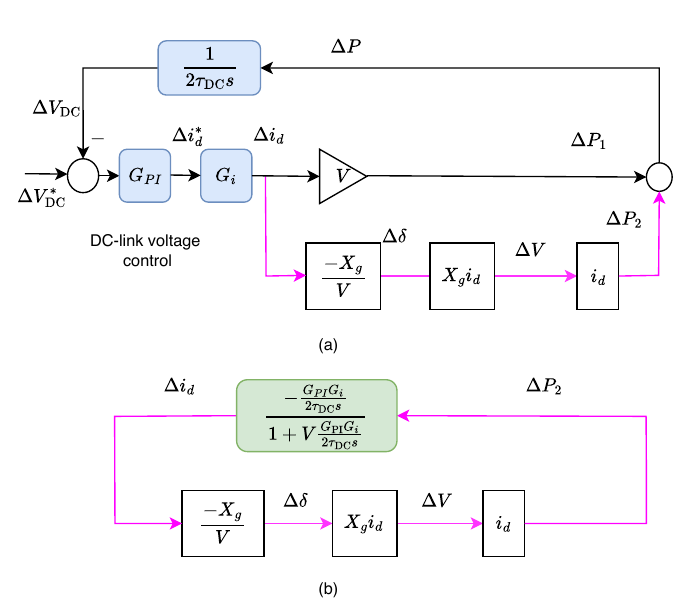}
\end{center}
\caption{Block diagrams. (a) The full diagram. (b) The block diagram with a part of it aggregated.  }
\label{fig:blockdiagram}
\end{figure}

The small-signal model of the above equation is as follows if $\delta$ is assumed as $0$ at the initial condition. 
\begin{align}
&\Delta V + jV \Delta \delta = -jX_g (\Delta i_d + ji_d \Delta \delta) \notag\\
\Longrightarrow & \Delta V = X_g i_d \Delta \delta, \> V \Delta \delta = -X_g \Delta i_d
\end{align}

Therefore, it can be seen that $\Delta V$ can be influenced by $\Delta i_d$ in the following way:
\begin{align}
\Delta V = -\frac{X^2_g i_d}{V} \Delta i_d. 
 \label{eq:V}
\end{align}
If the real current demand from the load increases, the AC voltage will reduce. The real power consumed by the load is influenced by both the real current $i_d$ and the voltage $V$. Their relationship and the small-signal relationship are as follows:
\begin{align}
 P = Vi_d, \Longrightarrow \Delta P = V\Delta i_d + i_d \Delta V.
 \label{eq:P}
\end{align}

Eqs. \eqref{eq:V}, \eqref{eq:P} and the DC-link voltage control can be integrated into a feedback system, as shown in Fig. \ref{fig:blockdiagram}(a).
In Fig. \ref{fig:blockdiagram}(a), the lagging effect between $i_d$ and $i_d^*$ due to inner current control has been considered: \[i_d = \frac{1}{\tau_i s +1 } i^*_d.\] $G_{\rm DVC}$ is updated, as shown in \eqref{eq:G_DVC2}. 
\begin{align}
G_{\rm DVC} =\frac{\Delta V_{\rm dc}}{\Delta V^*_{\rm dc}}&= \frac{\displaystyle\frac{1}{\tau_i s+1 } \left(K_p + \frac{K_i}{s}\right)\frac{1}{2\tau_{\rm DC} s}}{1+ \displaystyle \frac{1}{\tau_i s+1 } \left(K_p + \frac{K_i}{s}\right)\frac{1}{2\tau_{\rm DC} s}} \notag\\
&=\frac{\displaystyle K_p s + K_i}{\displaystyle 2\tau_{\rm DC}\tau_i s^3 + 2\tau_{\rm DC}s^2 + K_p s+ K_i}.
\label{eq:G_DVC2}
\end{align}

It can be seen from Fig. \ref{fig:blockdiagram}(b) that a feedback system has been formed, with the operating condition and grid strength information included. The  subsystem from $\Delta P_2$ to $\Delta i_d$ has the transfer function of $-G_{\rm DVC}$ (assuming $V=1$), or the same as the closed-loop DVC system with a negative sign. In the end, the loop gain of the feedback system (based on the default negative feedback assumption) is: 
\begin{align}
\text{Loop gain} = -G_{\rm DVC} \, (X_g i_d)^2
\end{align}

%

Additionally, the synchronizing effect between the current and the voltage may be viewed as not immediate, rather with lagging. In Fig. \ref{fig:circuit}, the current and voltage are assumed to have the same phase angle. The underlying assumption is that the time for synchronizing has been omitted. In the revised diagram, this effect may be considered as another unit similar to a low-pass filter (LPF):  \[\Delta \delta= G_{\rm sync} \,\Delta \delta_{V},\] where $\delta$ is the synchronizing angle and $\delta_v$ is the voltage's phase angle.  

The KVL may be rewritten as: 
\begin{align}
Ve^{j\delta_V} = \overline{V}_g -jX_g\cdot i_d e^{j\delta}. 
\end{align}
The small-signal model of the above equation is as follows if $\delta_V=\delta$ is assumed as $0$ at the initial condition. 
\begin{align}
&\Delta V + jV \Delta \delta_V = -jX_g (\Delta i_d + ji_d \Delta \delta) \notag\\
\Longrightarrow & \Delta V = X_g i_d \Delta \delta, \>
 \Delta \delta_V = -X_g \Delta i_d
\end{align}

\begin{figure}[!ht]
\begin{center}
\includegraphics[width=3.5in]{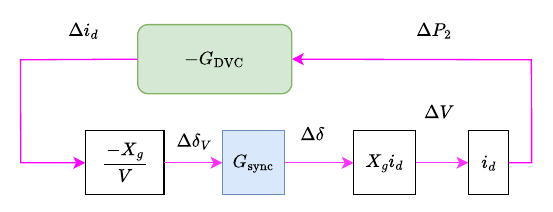}
\end{center}
\caption{Block diagram. }
\label{fig:blockdiagram2}
\end{figure}

The block diagram in Fig. \ref{fig:blockdiagram}(b) is now updated, as shown in Fig. \ref{fig:blockdiagram2}, where $\Delta i_d$ influences $\Delta \delta_V$, which further influences $\Delta \delta$ through a synchronizing unit $G_{\rm sync}$. 

It can be seen that $\Delta V$ can be influenced by $\Delta i_d$ in the following way:
\begin{align}
\Delta V = -X_g^2 i_d \,G_{\rm sync}\,\Delta i_d. 
 \label{eq:V1}
\end{align}

The loop gain becomes:
\begin{align}
\text{Loop gain} = -G_{\rm DVC} \, G_{\rm sync}\, (X_g i_d)^2
\label{eq:loopgain}
\end{align}
The addition of another low-pass filter $G_{\rm sync}$ leads to phase shifting (from $-180^{\circ}$ to $180^{\circ}$) at a frequency with a close proximity at the resonance frequency of $G_{\rm DVC}$. 

\begin{figure}[!ht]
\centering
\subfloat[]{\includegraphics[width=3.1in]{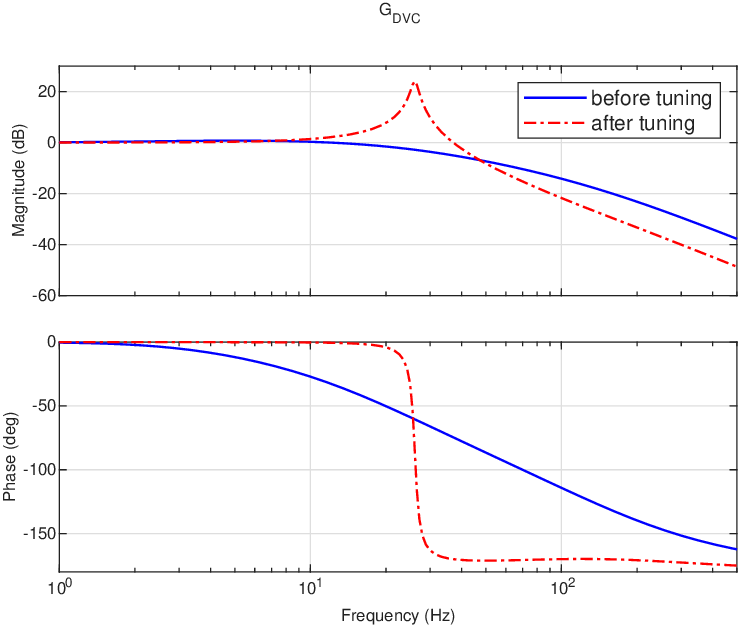}}\\
\subfloat[]{\includegraphics[width=3.1in]{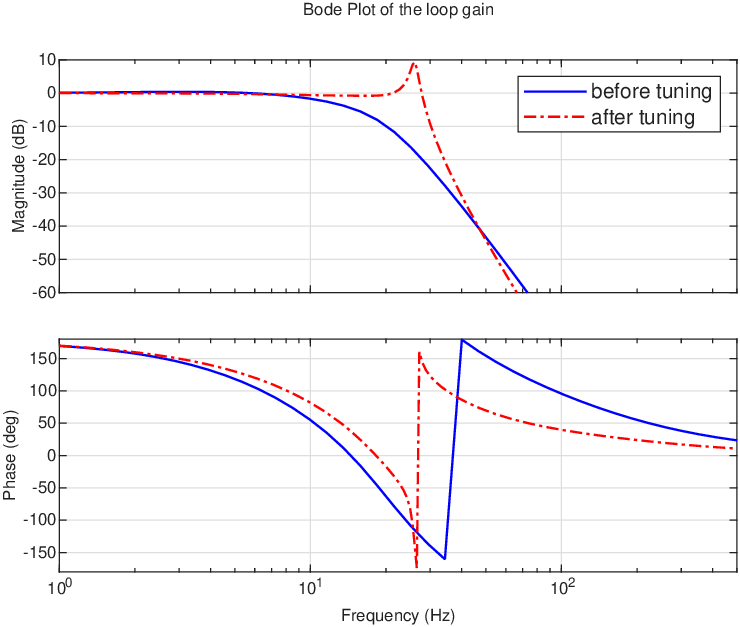}}\\
\caption{(a) Bode diagrams of the DVC closed-loop system. (b) Bode diagrams of the loop gain.   }
\label{fig:case_1}
\end{figure}
\begin{figure}[!ht]
\centering
\subfloat[]{\includegraphics[width=3.2in]{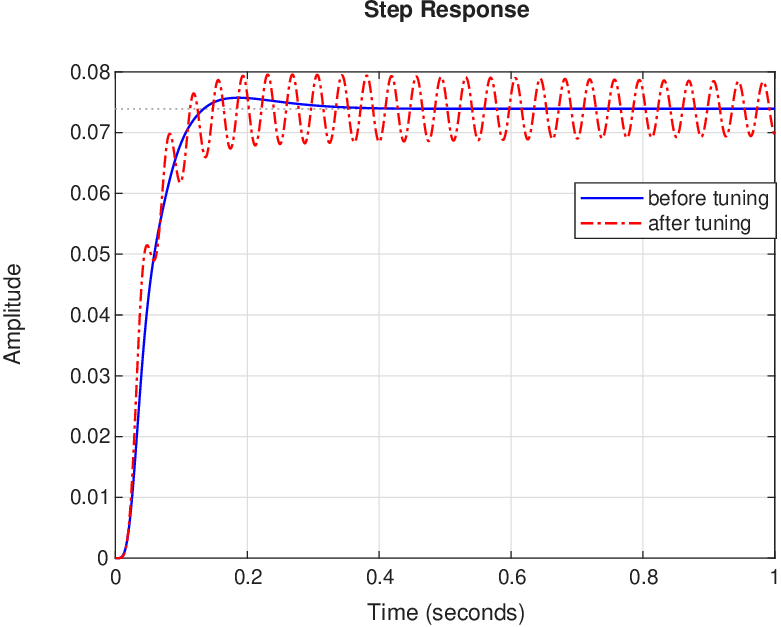}}\\
\subfloat[]{\includegraphics[width=3.2in]{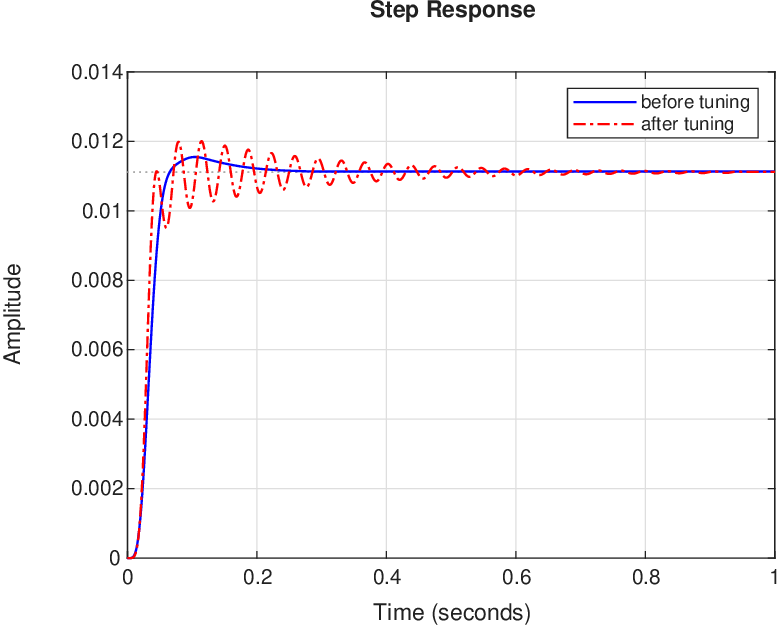}}
\caption{Step responses of the closed-loop system. (a) $(i_dX_g)^2 =0.425$. (b) $(i_dX_g)^2 = 0.1$.   }
\label{fig:step}
\end{figure}

\subsection{Numerical case}
A numerical case is presented to demonstrate the creation of $\sim23$-Hz oscillations. Two sets of the PI control parameters have been used and they are named as ``before tuning'' and ``after tuning''. The related control parameters are presented in Table \ref{tab1}.

\begin{table}[htp]
\footnotesize
\caption{Parameters of the feedback system}
\begin{center}
\begin{tabular}{c|c|c}
\hline \hline 
  Description  &  Parameters & Comments \\
    \hline
DVC PI controller & $10+ \frac{1}{0.0063s}$ & before tuning\\
\hline
  DVC PI controller & $2.8+\frac{1}{0.0005s}$ & after tuning \\
\hline
DC-link time constant $\tau_{\rm DC}$ & 0.0377 s & pu \\
\hline
$G_i$ & $\frac{1}{1+0.001s}$ & \\ \hline
$G_{\rm sync}$ & LPF    &  subject to tuning  \\ \hline
$i_d $ & 1 & pu \\ \hline
$X_g$ &  $0.65$ & pu \\
\hline \hline
\end{tabular}
\end{center}
\label{tab1}
\end{table}%

Fig. \ref{fig:case_1}(a) shows the Bode plots of $G_{\rm DVC}$.  The DVC performance is similar to a low-pass filter. 
The parameters of the DC-link voltage control have been turned to create a resonance peak at $26$ Hz. A typical parameter of $\tau_{\rm DC}$ of $0.0377$ s is used. This parameter is comparable to the DC-link capacitor's time constant $0.0272$ s for a 2-MW type-4 wind turbine. The wind turbine's DC-link capacitor size ($0.09$F) and the nominal DC-link voltage $1100$ V \cite{fan2018wind}.  

The current tracking performance is assumed as a first-order LPF: $1/(0.001s+1)$. The PI controller of the DC-link voltage is selected as $(2.8, 2000)$ to achieve a bandwidth above $30$ Hz and a resonant peak at $26$ Hz. At the proximity of the resonant frequency of $25$ Hz, the angel changes from $0^{\circ}$ to $-180^{\circ}$, while at the resonant frequency, the angle is $-90^{\circ}$.

Fig. \ref{fig:case_1}(b) shows the loop gain when $i_dX_g =1$, or $-G_{\rm DVC}G_{\rm sync}$, which is related to the open-loop gain in \eqref{eq:loopgain}.  With the additional phase lag introduced by the synchronizing unit $G_{\rm sync}$, $-G_{\rm DVC}G_{\rm sync}$ experiences phase shifting (the phase angle changes from $-180^{\circ}$ to $180^{\circ}$ at the proximity of the resonant frequency. If the proportional gain of the loop gain $(X_g i_d)^2$ makes the magnitude of the loop gain exceeds $0$ dB at the phase shifting frequency, the system will experience sustained oscillations at $26$ Hz. 

Fig. \ref{fig:step}(a) shows the step responses of the closed-loop systems based on the two sets of DVC parameters, and with $(X_g i_d)^2$ at $0.425$. This closed-loop system has its input as the power imbalance and output as the AC voltage. The system is subject to $0.1$ pu sudden power unbalance, due to sudden drop of DC load. In turn, the DC-link voltage increases and the real current $i_d$ decreases due to the DVC effect. A decrease in $i_d$ further leads to an increase in the AC voltage. Therefore, the AC voltage increases.  The tuned parameters  lead to $26$-Hz oscillations upon the step change, while the other set of parameters do not lead to any oscillations.

When the operating condition changes, e.g., real power exporting level reduces and $(X_gi_d)^2=0.1$, the closed-loop system will not experience sustained oscillations.  Fig. \ref{fig:step}(b) shows the step responses of the closed-loop systems when $(X_g i_d)^2$ at $0.1$. Oscillations are damped. 

\emph{Remarks:} This numerical case demonstrates two aspects of oscillations related to LELs. 
\begin{itemize} \item Oscillations may be generated when a LEL's power consumption level is increased to a certain level. 
\item  The oscillation frequency has to do with the DC-link voltage control and the synchronizing unit. Their bandwidths and resonant frequency determine the oscillation frequency. \end{itemize}


\section{EMT testbed construction and simulation results}
\subsection{EMT testbed circuit and control details}
Besides the frequency-domain analysis, an electromagnetic transient (EMT) simulation testbed has been constructed to validate the analytical results. The EMT testbed was built in PSCAD.
\begin{figure}[!ht]
\begin{center}
\includegraphics[width=3.5in]{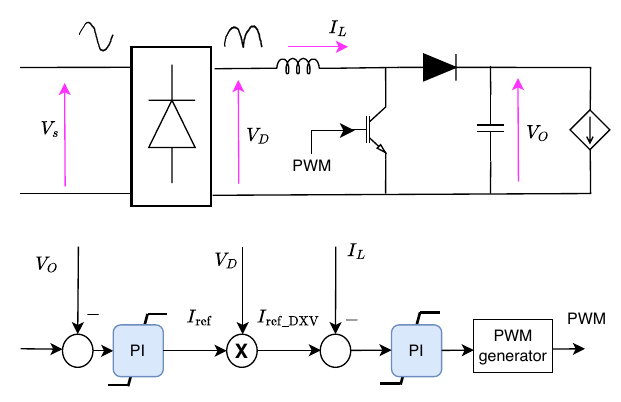}
\end{center}
\caption{Power electronic circuit and the PFC control scheme for a 1-MW load. Circuit parameters: 
$L = 10 \> \mu$H and $C=417$ mF. The DC-link voltage controller is: $10+1/(0.0063s)$ and the inner current controller is $30+1/(0.00025s)$. }
\label{fig:PFC}
\end{figure}

\begin{figure}[!ht]
\begin{center}
\includegraphics[width=2.80in]{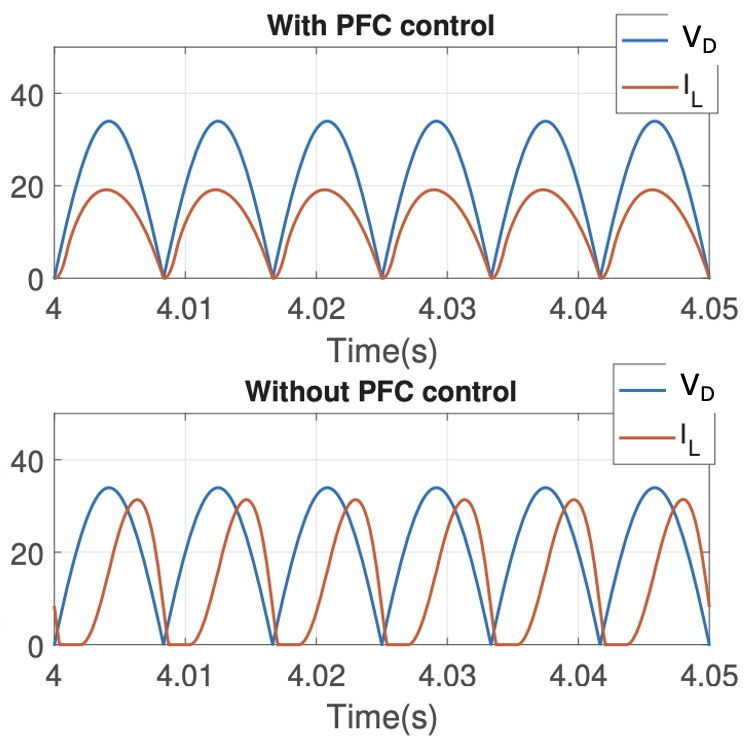}
\end{center}
\caption{Inductor current and rectified voltage with and without the PFC control scheme. }
\label{fig:PFC1}
\end{figure}
This load device-level model of a 1-MW single-phase cryptocurrency miner, developed jointly by ERCOT and Texas A\&M University (TAMU), is derived from a commercially available mining unit \cite{ERCOT_PSCAD, samanta2024EMT}. The model has been validated against laboratory test data to ensure accurate representation of its dynamic and steady-state behavior \cite{ERCOT_PSCAD, samanta2024EMT}. Fig. \ref{fig:PFC} shows the power electronic circuit at the device level. As shown in the figure, the power conversion process consists of multiple stages, beginning with a diode bridge for AC–DC conversion, followed by subsequent DC-link and load interface components. The input AC voltage has a nominal RMS voltage of 240 V.  The resulting voltage right after the rectifier $V_D$ has an average value of 216 V. The DC-DC boost converter boosts 216 V to 425 V, and the nominal voltage of the DC-link is 425 V, while the DC-link capacitor is $417$ mF for a 1-MW unit. These parameters lead to a time constant $\tau_{\rm DC}$ at  $0.0377$ s. Furthermore, the DC current has a nominal value of 2.5 kA.

The controller of the PSU, also shown in Fig. \ref{fig:PFC}, has  two functions: DC-link voltage regulation and current reference tracking. 
{The DC-link capacitor's voltage $V_o$ is measured and compared with its reference value. This error is amplified by a PI controller to generate a current reference $I_{\rm ref}$. At steady state, this reference is constant. On the other hand, the current $I_L$ right after the rectifier is to be regulated and this DC current has harmonics. To be synchronized with the grid, the current reference is generated to be in phase with the rectified voltage $V_D$.  Fig. \ref{fig:PFC1} from the prior work of the team members of this project \cite{bao2021modeling} illustrates the effect of PFC control.  }
A proportional integral (PI) controller is deployed to ensure that the current tracks the current reference.

\begin{figure*}[!ht]
\begin{center}
\includegraphics[width=6.5in]{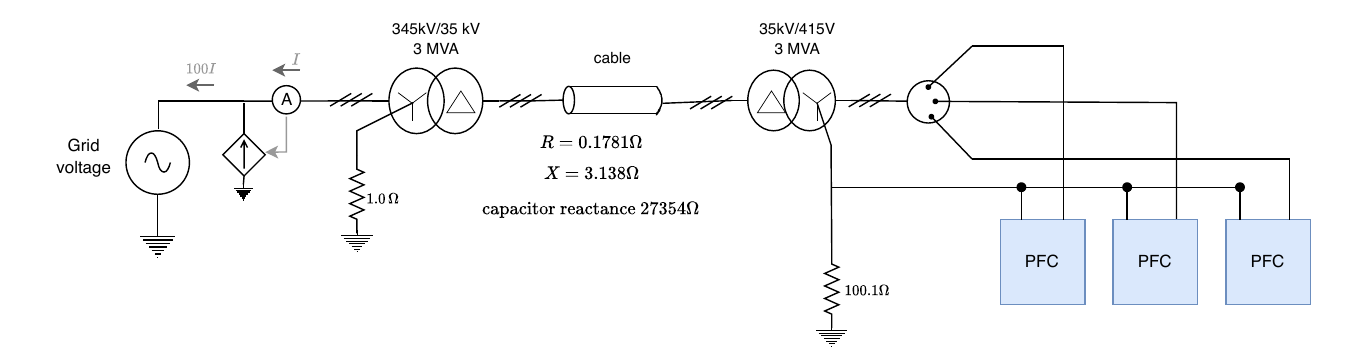}
\end{center}
\caption{The EMT testbed circuit topology for a grid-integrated 300-MW load. }
\label{fig4}
\end{figure*}
\begin{figure}[!ht]
\begin{center}
\includegraphics[width=3.5in]{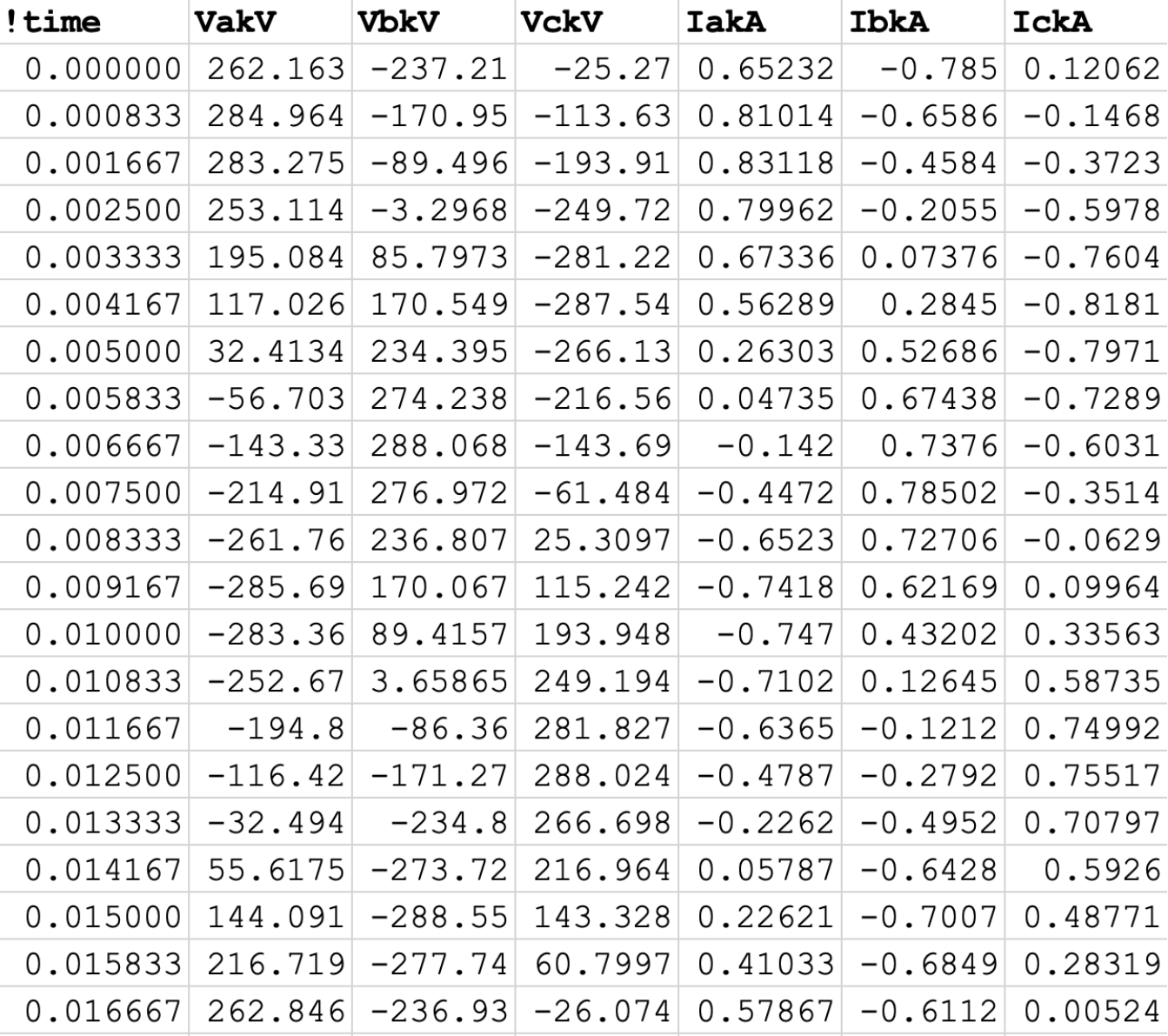}
\end{center}
\caption{Real-world DFR data. }
\label{fig5}
\end{figure}

Compared to the model for analysis, the simulation model has the additional details of the PFC or the current control details. In the analytical model, the LEL is treated as a controllable current source. On the other hand, in the simulation testbed, the LEL is treated as a PSU controlled DC load, which is more aligned to the physical system. 


The PSCAD testbed of a 300 MW load is shown in Fig. \ref{fig4}. The 3-MW three-phase load has been scaled up by 100 times to represent the LEL of 300 MW. The grid voltage is at 345 kV. Two step-down transformers reduce the AC voltage level to 415 V, or 240 V per phase. At each phase, a power electronic circuit shown in Fig. \ref{fig:PFC} is connected to the DC loads.  
Since detailed transformer and collector system information were not available, the facility model was constructed using typical and industry-standard assumed parameters. This approach allows for a reasonable representation of the aggregate behavior of a large-scale crypto mining facility under various grid conditions.

The Digital Fault Recorder (DFR) data, 20 samples per cycle, are shown in Fig. \ref{fig5}. The data are for the instantaneous phase voltages and phase currents. This data is played back in PSCAD (fed to the three-phase grid voltage) to evaluate the response of the facility model developed. It  was used to replicate the oscillations after a sudden power rise event for the crypto miner facility.  

\subsection{Parameter tuning}

\begin{figure}[!ht]
\begin{center}
\includegraphics[width=3.5in]{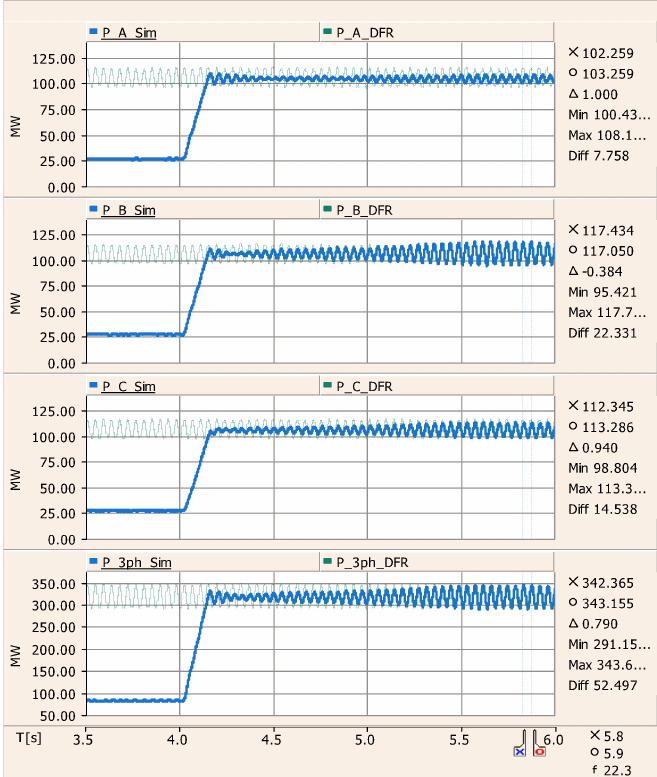}
\end{center}
\caption{EMT testbed simulation results of real power measured at each phase and the total real power of three phases. The load's power ramps up to 320 MW. Oscillations of $22.3$ Hz appear. 
}
\label{fig:sim1}
\end{figure}

\begin{figure}[!ht]
\begin{center}
\includegraphics[width=3.5in]{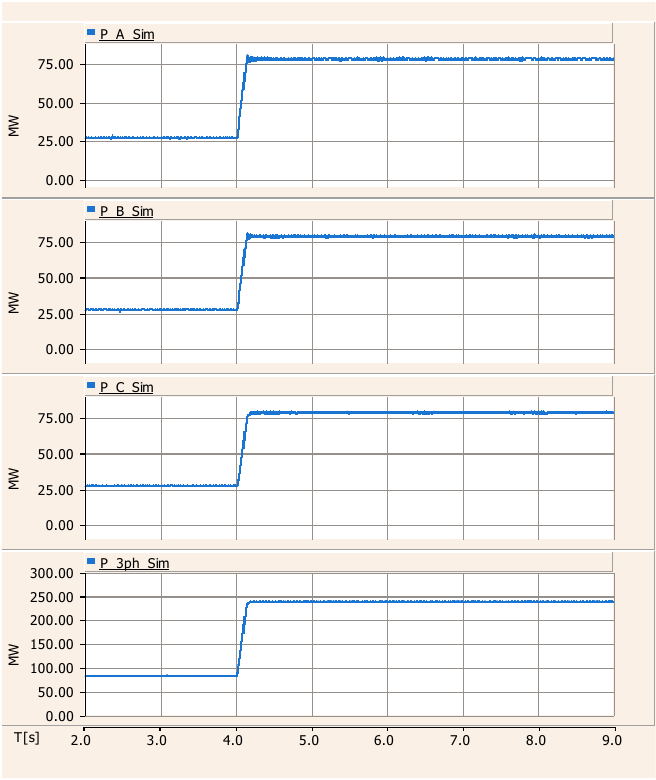}
\end{center}
\caption{EMT testbed simulation results. The load's power ramps up to 240 MW. There are no oscillations in real power.  }
\label{fig:sim2}
\end{figure}

\begin{figure}[!ht]
\begin{center}
\includegraphics[width=3.5in]{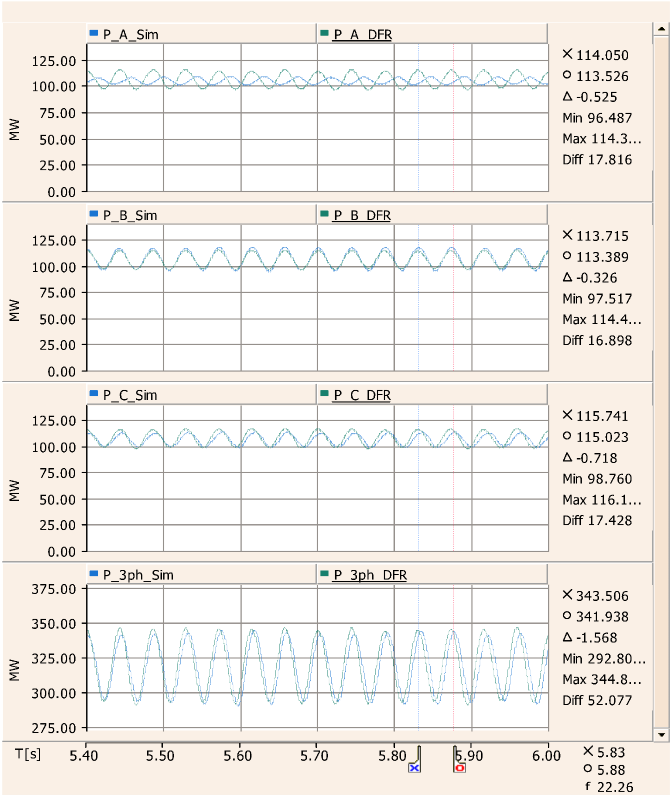}
\end{center}
\caption{EMT testbed simulation results. Zoom in of Fig. \ref{fig:sim1}.}
\label{fig:sim3}
\end{figure}

\begin{figure}[!ht]
\begin{center}
\includegraphics[width=3.5in]{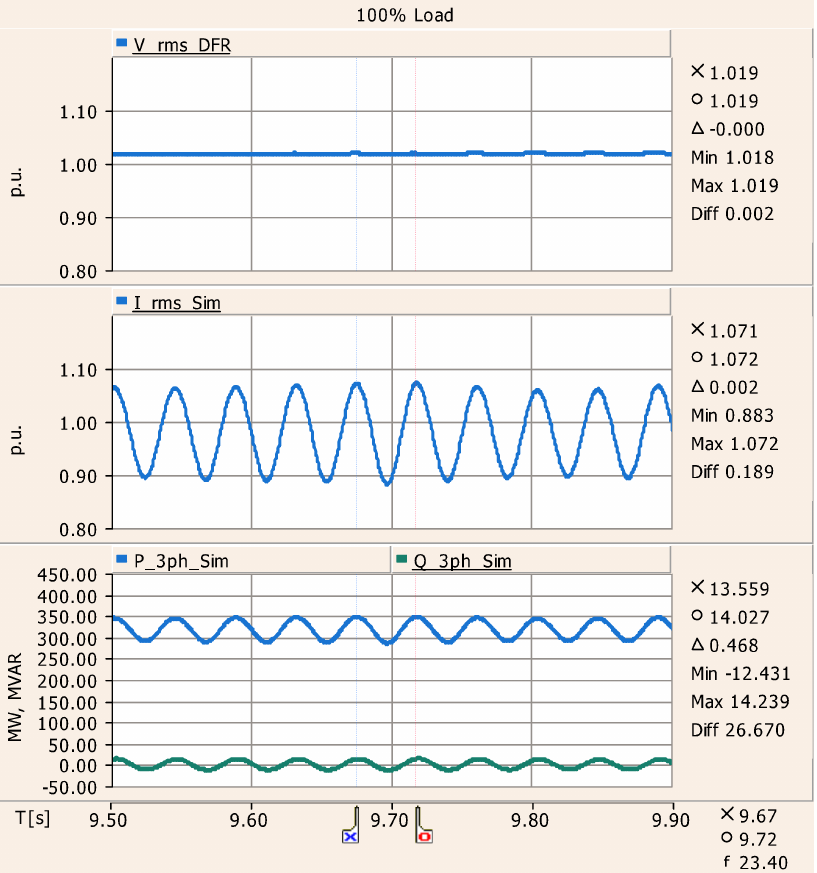}
\end{center}
\caption{EMT testbed simulation results: the AC grid voltage, AC current, real and reactive power.}
\label{fig:sim_ac}
\end{figure}

\begin{figure}[!ht]
\begin{center}
\subfloat[]{\includegraphics[width=3.5in]{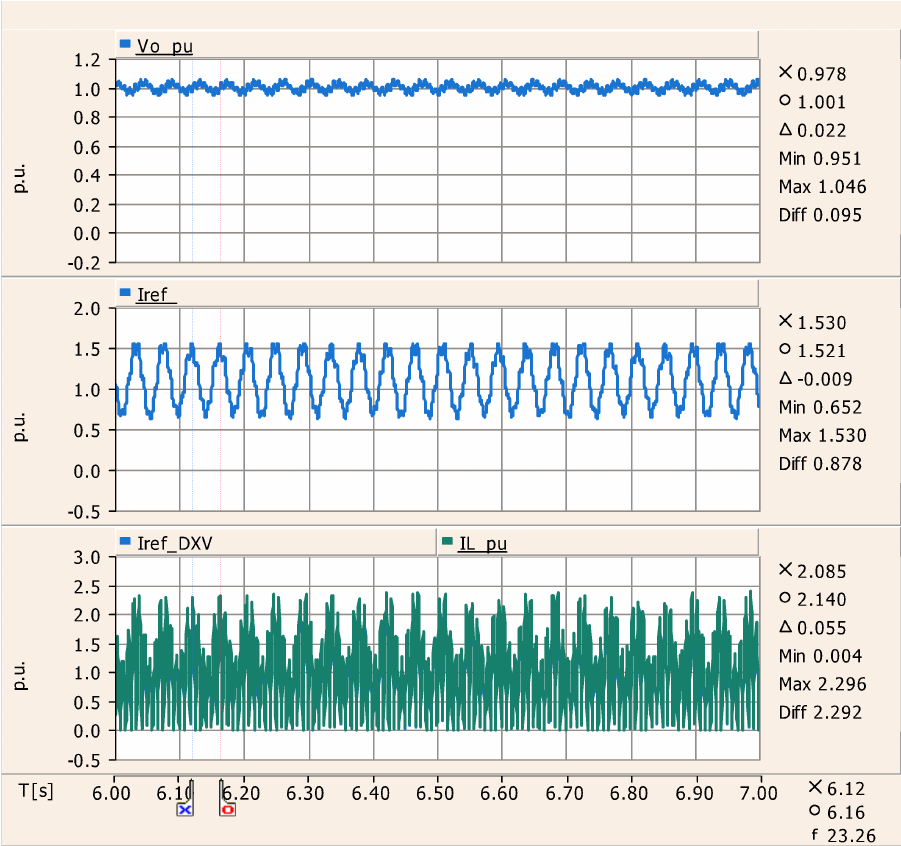}}\\
\subfloat[]{\includegraphics[width=3.5in]{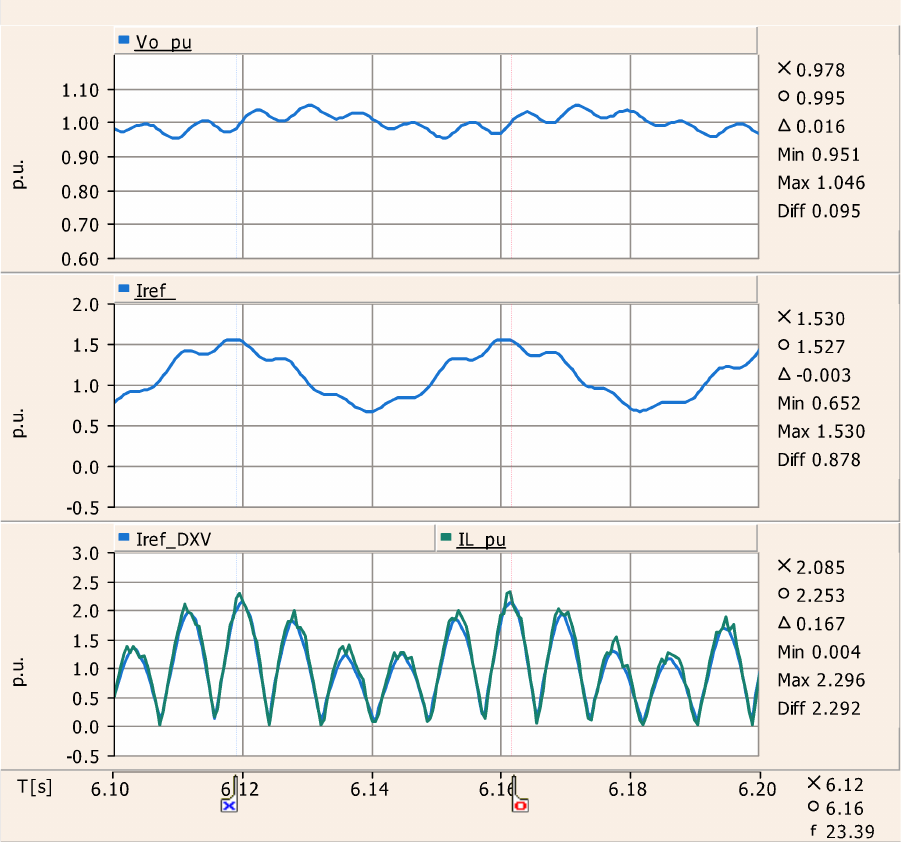}}
\end{center}
\caption{EMT testbed simulation results. (a) The DC-link capacitor voltage $V_o$, the output of the DVC controller $I_{\rm ref}$, the reference and the  current through the inductor. (b) Zoom-in of (a).}
\label{fig:sim4}
\end{figure}

The DC voltage controller parameters were intentionally tuned to reproduce the oscillatory response observed in the field. 
The tuned DC voltage controller parameters are: proportional gain = 2.8 (it was 10),  integral time constant = 0.0005 (it was 0.0063). 

Analysis has been carried to examine the DVC control resonant frequency in Section II. Using the parameters of $K_P = 2.8$, $K_i=2000$ or $1/0.0005$, and $\tau_{\rm DC} = 0.0377$ s, the resonant frequency of the closed-loop DVC control is about $26$ Hz, according to \eqref{eq:G_DVC2}, when the current tracking time is assumed to be $0.001$ s. It is to be noted that \eqref{eq:G_DVC2} provides an estimated resonant frequency for the DVC system. For instance, the current tracking system as a first-order lagging system is an estimation. On the other hand, the estimated resonant frequency is considered to be quite close to $23$ Hz. 

\subsection{EMT simulation results}
In Fig. \ref{fig:sim1}, the load's power increased to 100\% level (320 MW) at 4 s. As can be seen, the oscillations were observed. In the subplots, both active power (MW) waveforms from the DFR data and simulation results were overlayed to benchmark the model’s dynamic behavior in replicating real-world event oscillations. 

In Fig. \ref{fig:sim2}, the load's power increased  to 75\% level (240 MW) at 4 s. As can be seen, the real power  shows no oscillations. These two tests indicate that the instability mechanism is dependent on the operating conditions.

Fig. \ref{fig:sim3} presents a zoom-in view of the simulation results in Fig. \ref{fig:sim1}, highlighting the oscillatory behavior of the crypto miner facility. A $50$ MW peak-to-peak oscillation was observed at an approximate frequency of $23$ Hz, closely replicating the characteristics of the real-world event recorded in the field, shown in Fig. 2.

Fig. \ref{fig:sim_ac} presents the simulation results of the AC side voltage RMS, current RMS, real and reactive power. The first subplot shows the grid voltage profile based on the DFR data. The second subplot shows the RMS current into the load. The third subplot shows the real and reactive power into the load. It can be seen that $23$-Hz oscillations appear in the current RMS measurement, with a peak-peak magnitude close to $0.19$ pu. 
On the other hand, the grid voltage remains essentially flat, exhibiting negligible oscillations. This observation indicates that the 23-Hz oscillations originate from the load dynamics. The grid is sufficiently strong and therefore does not exhibit noticeable voltage oscillations, even when the load absorbs oscillatory power from the grid.
Comparing the real and reactive power, it can be seen that the oscillations are more observable in the real power, implying that real power and real current are the major influencing factors of the oscillations. 

Fig. \ref{fig:sim4} shows the case when real power ramps to $320$ MW and the simulation results of the DC-link capacitor voltage $V_o$, the output of the DVC controller $I_{\rm ref}$, the reference current $I_{\rm ref\_DXV}$, and the current through the inductor. The zoom-in view is shown in Fig. \ref{fig:sim4}(b). 

 It can be seen that the DC-link voltage and the output of the DVC (the reference of the real current) both show $23.26$ Hz oscillations. The PFC ensures that the rectifier current is in phase with the voltage right after the rectifier. This in turn ensures that the AC current flowing into the rectifier is in phase with the AC input voltage to the rectifier.  Additionally,  the real current reference $I_{\rm ref}$ has a DC value of 1.072 pu, while the $23$-Hz oscillation has a peak-peak amplitude of 0.878 pu. In the DC-link voltage $V_o$, the oscillation amplitude is very small (0.095 pu). This observation shows that the DC-link voltage controller may amplify oscillations at the frequency close to its resonant frequency.

\emph{Remarks:} The EMT simulation results corroborate the analysis results that the DC-link voltage control in the PSU of loads may lead to oscillations. Higher power consumption makes oscillations worse. And the oscillation frequency is related to the DC-link voltage control's resonant frequency.  


\section{Conclusion}
This research focuses on mechanism analysis and EMT replication of a real-world oscillation event in Texas. This oscillation event is unprecedented  since it is caused by a power electronic interfaced large electronic load.  It is found that similar to inverter-based resources, loads' converter control may also interact with grid and lead to oscillations. This particular phenomenon is associated with the DC-link voltage control in the power supply units. Through its influence on real current, the real current's influence on AC voltage and real power, and in turn the real power's effect on the DC-link voltage, an oscillatory mechanism is created. In this paper, the inter-winded relationship has been characterized as a feedback system, which is convenient for stability analysis and oscillation root cause analysis. The  analysis results obtained from the feedback system were validated using the EMT simulation results based on a detailed model with power electronic converter switching and control dynamics. We have arrived at a reasonable speculation that the DC-link voltage control of the PSU employed in loads can lead to oscillations at the resonant frequency. High power consumption can make oscillations appear. 

\bibliographystyle{IEEEtran}
\bibliography{bib}

\end{document}